\documentclass[twocolumn,showpacs,preprintnumbers,amsmath,amssymb,showkeys,usletter,10pt,pre]{revtex4-1}
\usepackage[]{graphicx}
\usepackage[]{color}

\newcommand{\Dt}{\Delta t}
\newcommand{\Dx}{\Delta x}
\newcommand{\Dv}{\Delta v}
\renewcommand{\d}{\mbox{d}}
\newcommand{\dt}{\mbox{d}t}
\newcommand{\fc}{{f_{\rm c}}}
\newcommand{\fNyq}{{f_{\rm Nyq}}}
\newcommand{\fsample}{{f_{\rm sample}}}
\newcommand{\kT}{k_{\rm B} T}
\newcommand{\kB}{k_{\rm B}}

\newcommand{\Tmsr}{t_{\mbox{\scriptsize msr}}}

\newcommand{\nwin}{n_{\rm win}}
\newcommand{\Ftherm}{F_{\rm therm}}

\newcommand{\beq}{\begin{equation}}
\newcommand{\eeq}{\end{equation}}
\newcommand{\bea}{\begin{eqnarray}}
\newcommand{\eea}{\end{eqnarray}}
\newcommand{\e}{\enspace}
\newcommand{\la}{\langle}
\newcommand{\ra}{\rangle}
\newcommand{\LA}{\left<}
\newcommand{\RA}{\right>}

\begin{document}
\title{ 
Harmonic Oscillator in Heat Bath:
Exact simulation of time-lapse-recorded data,
exact analytical benchmark statistics}
\date{\today}

\author{Simon F. N\o{}rrelykke\\
 {\small Department of Molecular Biology, Princeton University, Princeton, New Jersey, USA.}\\
 and \\
 Henrik Flyvbjerg\\
 {\small Department of Micro- and Nanotechnology, Technical University of Denmark,
Kongens Lyngby, Denmark.}}

\begin{abstract}
The stochastic dynamics of the damped harmonic oscillator in a heat bath
is simulated with an algorithm that is exact for time steps of arbitrary size.
Exact analytical results are given for correlation functions and power spectra
in the form they acquire when computed from experimental time-lapse recordings.
Three applications are discussed:
(i) Effects of finite sampling-rate and -time, described exactly here,
are similar for other stochastic dynamical systems---e.g.~motile micro-organisms and their time-lapse recorded trajectories.
(ii) The same statistics is satisfied by any experimental system
 to the extent it is interpreted as a damped harmonic oscillator
 at finite temperature---such as an AFM cantilever.
(iii) Three other models of fundamental interest are limiting cases
of the damped harmonic oscillator at finite temperature;
it consequently bridges their differences and describes
effects of finite sampling rate and sampling time for these models as well.
Finally, we give a brief discussion of  nondimensionalization. \\
\end{abstract}

\keywords{Ornstein-Uhlenbeck process, power-spectral analysis, mean-squared displacement, discrete sampling, discretization effects, Langevin, Brownian motion, optical trap, leakage, non-dimensionalization, correlation function}

\maketitle
\section{Introduction}
The damped harmonic oscillator in a heat bath
is the archetypical bounded Brownian dynamical system with inertia
and the simplest possible of this kind.
Analytically solvable, it offers insights that are valid also for more complex systems.
Its experimental correlation functions and power spectra are given analytically here in the form they take when computed from time-lapse recorded trajectories.
Such trajectories are generated and analyzed for illustration in an exact Monte Carlo simulation.
The results differ significantly from those derived from the standard continuous-sampling formulation.

In mathematical terms, it is the Ornstein-Uhlenbeck (OU) process
in a Hookean force field we treat,
and the results discussed here extend the ones given in \cite{Norrelykke2010,SchimanskyGeier1990} as well as the free case treated by Gillespie~\cite{Gillespie1996}.
Three classical papers on the  OU-process are~\cite{Uhlenbeck1930,Chandrasekhar1943,Wang1945}.
Historically, the model was prompted by a question by Smoluchowski regarding how inertia might modify
Einstein's theory of Brownian motion~\cite{Ornstein1919}.
H. A. Lorentz soon after pointed out that Ornstein's answer to that question
was insufficient for the case of classical Brownian motion, i.e., in a fluid of density similar to
the Brownian particle~\cite{Lorentz1921}:
Hydrodynamical effects, entrainment and back-flow, were more important with the time resolution
that was available in the first century of Brownian motion.
Inertia in classical Brownian motion became experimentally relevant only with precision calibration of optical tweezers 
\cite{BergSorensen2004,BergSorensen2006}
and was directly observed only in 2005 \cite{Lukic2005,Selmeczi2007}.

The OU process remained and remains, however, an important model for Brownian motion dominated by inertia,
such as massive particles~\cite{Burnham2009,Burnham2010,Li2010} or AFM cantilevers~\cite{Sader1998} in air,
and for other kinds of persistent random motion, e.g.\ cell migration \cite{Gail1970,Hall1977,Euteneuer1984,Selmeczi2005,Selmeczi2008,Liang2010,Li2008,Codling2008} and commodity pricing \cite{Schwartz2000}.
Consequently, the treatment given here of its experimental statistics
for time-lapse recorded data should be useful in several ways:

(i) All experimental statistics contain effects of finite sampling rate,
finite sampling time, and finite statistics.
They do so to various degrees, but the effects are inherent in measurements.
We give exact analytical expressions for these effects, for the model treated here.
Statistics for more complex systems contain the same effects, qualitatively,
and quantitatively as well, by degrees we estimate.

(ii) The statistics we describe below
must be found for any experimental system
to the extent that system is interpreted as a damped harmonic oscillator
at finite temperature---e.g.~an AFM cantilever to be
calibrated by interpretation of its thermal power spectrum.

(iii) Three other models are limiting cases
of the model treated here:
\begin{enumerate}
\item
At \emph{vanishing mass}, Einstein's theory for Brownian motion
in a harmonic trap, which, e.g., is a minimal model for the Brownian dynamics of
a microsphere held in an optical trap, with magnetic tweezers \cite{Velthuis2010}, or surface-tethered by DNA \cite{Beausang2007}.
\item
At \emph{vanishing external force},
the Ornstein-Uhlenbeck model of free Brownian motion with inertia,
which, e.g., is a minimal model for the persistent random motion seen in
trajectories of motile cells.
\item
At \emph{vanishing mass and external force},
Einstein's original theory for Brownian motion in a fluid at rest.
\end{enumerate}
The results for the harmonic oscillator, given below, carry over to these three models.
The limits are not all obvious, but always enlightening, hence described below. 

(iv) As the model treated here bridges the differences
between the three limiting cases,
the material presented here is well suited
for a pedagogical, hands-on computer-based introduction
to the four dynamic systems covered here:
free/bound diffusion, with/without inertia,
their equilibrium behavior, correlations, and power spectra,
and their transient behavior to equilibrium.\\

\section{Exact Discretized Einstein-Ornstein-Uhlenbeck theory of Brownian motion in Hookean force field}
\label{sec:EOUH}
The Einstein-Ornstein-Uhlenbeck theory for the Brownian motion of a damped harmonic oscillator in one dimension
is simply Newton's Second Law for the oscillator with a thermal driving force, a.k.a.\ the Langevin equation
for this system,
\bea  \label{eq:Newton}
	m\ddot{x}(t) + \gamma \dot{x}(t) + \kappa x(t) &=& \Ftherm(t) \\
	\Ftherm(t) &=& (2\kT \gamma)^{1/2} \eta(t) \label{eq:Newton2} \enspace.
\eea
Here $x(t)$ is the coordinate of the oscillator as function of time $t$,
$m$ its inertial mass,
$\gamma$ its friction coefficient,
$\kappa$ is Hooke's constant,
and $\Ftherm$ is the thermal force on the oscillator.
Equation~(\ref{eq:Newton2}) gives the amplitude of this thermal noise explicitly in terms of $\gamma$, the Boltzmann energy $\kB T$, and $\eta(t)$, which is a normalized white-noise process,
i.e., the time derivative of a Wiener process, $\eta=dW/dt$, hence
\beq  \label{eq:whitenoise}
	\langle \eta(t) \rangle = 0~~;
	~~\langle \eta(t) \eta(t)\rangle =\delta(t-t')  ~~\mbox{  for all } t, t' \enspace.
\eeq
Equation~(\ref{eq:Newton}) can be rewritten as two coupled first-order differential equations,
\beq
	\frac{\d}{\dt} \left( \begin{array}{c}  x(t) \\ v(t)  \end{array} \right) =
	-  \mathbf{M}   \left( \begin{array}{c}  x(t) \\ v(t)  \end{array} \right)
 	+ \frac{\sqrt{2D}}{\tau}  \left( \begin{array}{c}  0 \\   \eta(t)  \end{array} \right)
	\enspace,
	\label{eq:OUtwo}
\eeq
where we have introduced Einstein's relation $D=\kT/\gamma$ and the $2\times 2$ matrix
\beq
	\mathbf{M} = \left( \begin{array}{cc}  0 & -1 \\ \frac{\kappa}{m}  &  \frac{\gamma}{m} 	\end{array} \right)
 	= \left( \begin{array}{cc}  0 & -1 \\ \omega_0^2  &  \frac{1}{\tau} 	\end{array} \right)
 	\enspace.
\eeq
Here  $\omega_0=\sqrt{\kappa/m}$ is the cyclic frequency of the undamped oscillator,
and $\tau=m/\gamma$ is the characteristic time of the exponential decrease with time that the momentum
of the particle undergoes in the absence of all but friction forces.
Below, we shall also need the cyclic frequency
of the damped oscillator, $\omega=\sqrt{\kappa/m-\gamma^2/(4m^2)}= \sqrt{\omega_0^2 - 1/(4\tau^2)}$,
which is real for less than critical damping, $\gamma^2 < 4m\kappa $.

Equation~(\ref{eq:OUtwo}) is solved by
\beq
 	\left( \begin{array}{c}  x(t) \\ v(t)  \end{array} \right) =
	\frac{\sqrt{2D}}{\tau}  \int_{-\infty}^t  \d t' \, e^{- \mathbf{M}(t-t')}
 	\left( \begin{array}{c}  0 \\   \eta(t')  \end{array} \right)
	\enspace,
	\label{eq:OUtwosoln}
\eeq
which, for  arbitrary positive $\Delta t$, and with $t_j = j \Delta t$, $x_j=x(t_j)$, and $v_j=v(t_j)$, gives us the recursive relation
\beq  \label{eq:OUiter}
 	\left( \begin{array}{c}  x_{j+1} \\ v_{j+1}  \end{array} \right) =
 	e^{- \mathbf{M}\Delta t}   \left( \begin{array}{c}  x_{j} \\ v_{j}  \end{array} \right)
	+  \left( \begin{array}{c}  \Delta x_j \\ \Delta v_j    \end{array} \right)
	\enspace,
\eeq
where
\beq \label{eq:DxDv}
	\left( \begin{array}{c}  \Delta x_j \\ \Delta v_j  \end{array} \right) =
 	\frac{ \sqrt{2D} }{ \tau } 
	 \int_{t_j}^{t_j+\Dt}   \d t' \, e^{- \mathbf{M}(t_j + \Dt -t')}
 	\left( \begin{array}{c}  0 \\   \eta(t')  \end{array} \right)
\eeq
and the time-independent matrix exponential can be written
\beq \label{eq:expM}
	e^{ - \mathbf{M} \Delta t } = e^{ -\frac{\Dt}{2\tau} }
	\left[   \cos(\omega \Dt) \mathbf{I}
	+ \sin(\omega \Dt) \mathbf{J}   \right]
\eeq
with
\beq \label{eq:IJ}
	\mathbf{I} \equiv
	\left( \begin{array}{cc}
	1 & 0 \\[0.5ex]
	0 & 1 \end{array}\right) \mbox{and }
	\mathbf{J}\equiv
	\left( \begin{array}{cc}
	\frac{1}{2\omega\tau} & \frac{1}{\omega} \\[0.5ex]
	\frac{-\omega_0^2}{\omega} & \frac{-1}{2\omega\tau} \end{array} \right) \e.
\eeq
That the matrix exponential can be written this way,  can be proven in several ways. 
We used the algebra of Pauli matrices.  
Alternatively, one may observe that the cosine and sine terms on the RHS are the even and odd parts of the LHS\@.
The latter are straight-forwardly, if tediously, computed from the Taylor series for the exponential.  
Inserting Eq.~(\ref{eq:expM})  in Eq.~(\ref{eq:DxDv}) we see that
\bea \label{eq:Dx}
	\Dx_j &=& \frac{\sqrt{2D}}{\omega\tau} \int_{t_j}^{t_{j+1}} \!\!\!\!\!\!\!\!\! \dt \,
	e^{-\frac{t_{j+1}-t}{2\tau}} \sin(\omega(t_{j+1}-t)) \, \eta(t) \\
	\nonumber\\
	\nonumber\\
	 \label{eq:Dv}
	\Dv_j &=& - \frac{\sqrt{2D}}{2\omega\tau^2} \int_{t_j}^{t_{j+1}} \!\!\!\!\!\!\!\!\! \dt \,
	e^{-\frac{t_{j+1}-t}{2\tau}} \sin(\omega(t_{j+1}-t)) \, \eta(t) \nonumber \\
	&&\\
	&+& \frac{\sqrt{2D}}{\tau} \int_{t_j}^{t_{j+1}} \!\!\!\!\!\!\!\!\! \dt \,
	e^{-\frac{t_{j+1}-t}{2\tau}} \cos(\omega(t_{j+1}-t)) \, \eta(t)
	 \nonumber
\eea
are two correlated random numbers from zero-mean Gaussian distributions.
They can be written as a linear combination of two independent Gaussian variables:
The four parameters that determine this linear combination can be chosen at will, as long as the combination has the same variance-covariance as the two original correlated variables.
This is a direct consequence of the Gaussian distribution being completely determined by its mean and variance-covariance.
Thus, we can write
\bea \label{eq:Dx2}
	\Dx_j &=&  \sigma_{xx}  \,\, \xi_j \\
	\label{eq:Dv2}
	\Dv_j &=& \sigma^2_{xv} / \sigma_{xx}  \,\, \xi_j +
			\sqrt{ \sigma^2_{vv} - \sigma^4_{xv} / \sigma^2_{xx} } \,\, \zeta_j
	\enspace,
\eea	
where the $\sigma$s are elements of the variance-covariance matrix (see below),
and $\xi$ and $\zeta$ are independent random numbers with Gaussian distribution, unit variance, and zero mean.
This particular choice of linear combination mirrors the structure of Eqs.~(\ref{eq:Dx})~and~(\ref{eq:Dv}).
Using Eqs.~(\ref{eq:whitenoise}),~(\ref{eq:Dx}),~and~(\ref{eq:Dv}) we calculate that the elements of the variance-covariance matrix are, for $\omega \neq 0$
\bea
	\sigma^2_{xx} &\equiv& \la (\Dx_j)^2 \ra
	= \frac{D}{4 \omega^2 \omega_0^2 \tau^3} \left( 4\omega^2\tau^2 +\right. \label{eq:sxx} \\
	&& \left. e^{-\frac{\Dt}{\tau}} [ \cos( 2\omega\Dt ) - 2 \omega \tau \sin( 2\omega\Dt ) - 4\omega_0^2\tau^2] \right)
	\nonumber \\
	\nonumber \\
	\sigma^2_{vv} &\equiv& \la (\Dv_j)^2 \ra 
	= \frac{D}{4 \omega^2 \tau^3} \left( 4\omega^2\tau^2 +\right.  \label{eq:svv} \\
	&& \left. e^{-\frac{\Dt}{\tau}} [ \cos( 2\omega\Dt ) + 2 \omega \tau \sin( 2\omega\Dt ) - 4\omega_0^2\tau^2] \right)
	\nonumber \\
	\nonumber \\
	\sigma^2_{xv} &\equiv& \la \Dx_j \Dv_j \ra  
	= \frac{D}{\omega^2\tau^2} e^{-\frac{\Dt}{\tau}} \sin^2( \omega \Dt ) \label{eq:sxv} \enspace.
\eea
\\
At critical damping ($\omega=0$) this variance-covariance simplifies to
\bea
	\sigma^2_{xx} &\equiv& \la (\Dx_j)^2 \ra \label{eq:sxxcrit} \\
	&=& 4 D \tau \left( 1- e^{-\frac{\Dt}{\tau}} [ 1 + \Dt/\tau + \frac{1}{2} (\Dt/\tau)^2 ] \right)
	\nonumber \\
	\nonumber \\
	\sigma^2_{vv} &\equiv& \la (\Dv_j)^2 \ra \label{eq:svvcrit} \\
	&=&  D/\tau  \left( 1- e^{-\frac{\Dt}{\tau}} [ 1 - \Dt/\tau + \frac{1}{2} (\Dt/\tau)^2 ] \right)
	\nonumber \\
	\nonumber \\
	\sigma^2_{xv} &\equiv& \la \Dx_j \Dv_j \ra  
	= D e^{-\frac{\Dt}{\tau}} (\Dt/\tau)^2  \label{eq:sxvcrit} \enspace.
\eea

\begin{figure}
\includegraphics[width=\linewidth]{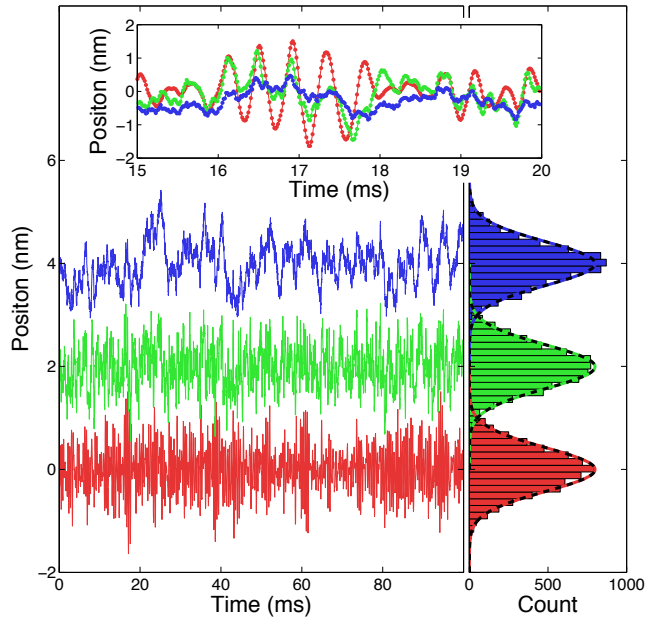}
\caption[]{\label{fig:AFM_positions}
Positions of harmonically trapped massive particles in a thermal bath.   
Three different regimes are studied by varying the drag coefficient above and below the critical value, where $\omega=0$, by an order of magnitude:
Underdamped (red, $\gamma = \gamma^{\rm crit} / 10$), critically damped (green, $\gamma = \gamma^{\rm crit}$, offset by 2\,nm), and overdamped (blue, $\gamma = 10\,\gamma^{\rm crit} $, offset by 4\,nm).
Simulation parameters in Eq.~(\ref{eq:OUiter}):  $m=1$\,ng, $\kappa=225$\,mg/s$^2$, $T=275$\,K,  $\fsample = 65,536$\,Hz, $(x(0),v(0))=(0,0)$, and the same $(\xi_j,\eta_j)$ are used for the three simulations.
The values for the underdamped case are alike to those for an AFM cantilever in water, whereas the critical and overdamped case corresponds to the same  cantilever in a more viscous environment or with a smaller spring constant. 
The insert shows a magnified portion of the trace, revealing the oscillating, critical, and random nature of the motion, respectively.
On the right, histograms show the distribution of the position data with $\kB T/\kappa$-variance Gaussians overlaid.}
\end{figure}

Figure~\ref{fig:AFM_positions} shows the simulated positions in the underdamped, critically damped, and overdamped regime.  
Note that all three regimes have the same mean (zero) and variance ($\kT/\kappa$).
Only at short times, as shown in the insert, does the difference between the three regimens reveal itself:  The 
position coordinate of the underdamped system oscillates, while the position coordinate of the overdamped system does not show discernible persistence of motion. The position coordinate of the critically damped system looks at times like it oscillates, but actually only displays positively correlated random motion, as revealed by  the velocity auto-correlation function given below and shown in Fig.~\ref{fig:AFM_corr}B.

\section{Mean-Squared-Displacement}
\begin{figure}
\includegraphics[width=\linewidth]{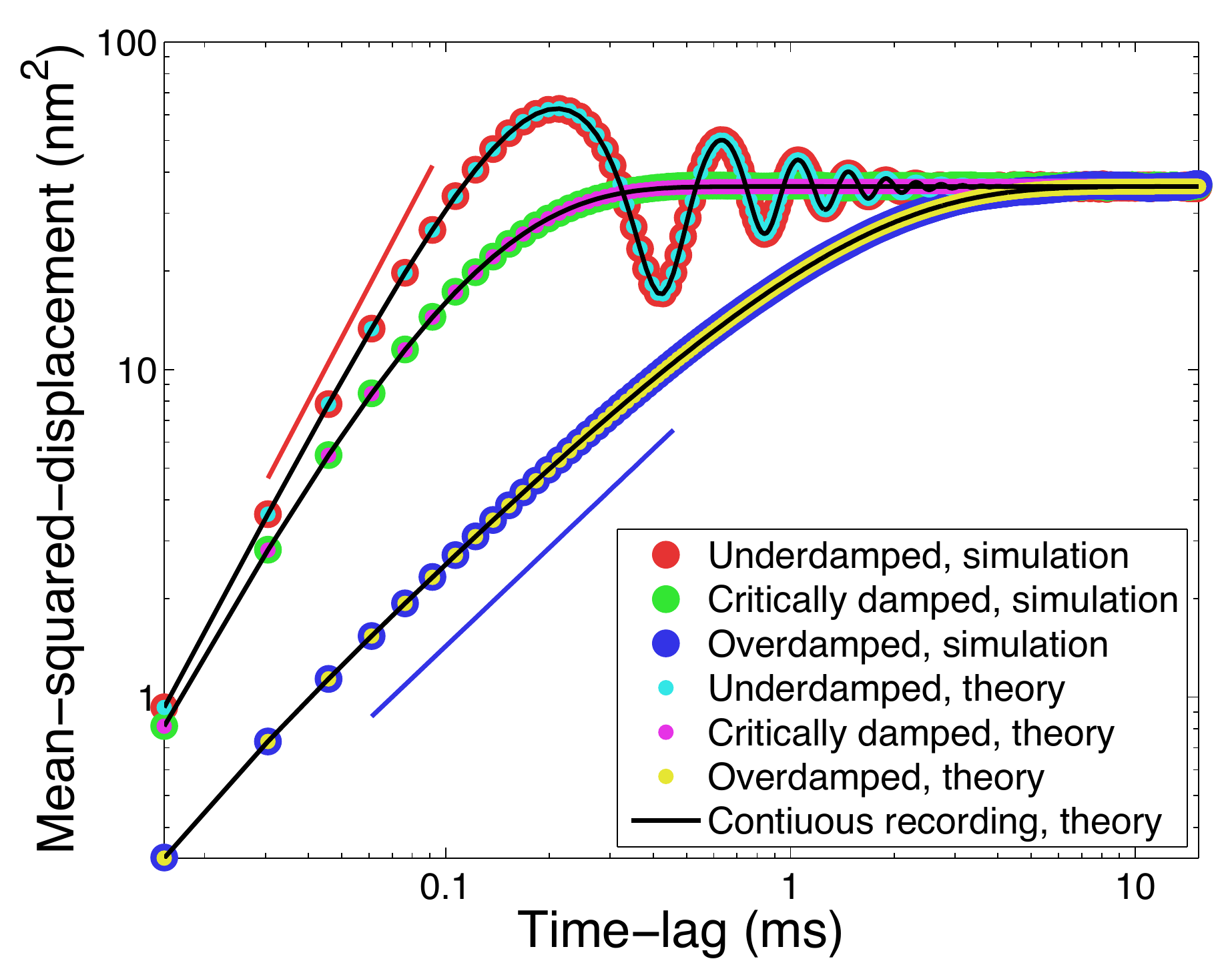}
\caption[]{\label{fig:AFM_MSD}
Mean-squared-displacements for harmonically trapped massive particles.  
Same simulation conditions as in Fig.~\ref{fig:AFM_positions}, except $\Tmsr=8$\,s.
The result of the stochastic simulations are shown in red, green, and blue.
Complementary colors (cyan, magenta, and yellow) show the MSD, Eq.~(\ref{eq:MSDcovar}), evaluated at discrete time-points with the same parameter values as in the simulation.
Thin black lines show the MSD Eq.~(\ref{eq:MSDcovar}) for  continuous time.
Red and blue lines show slopes of two and one, respectively.
Notice how the MSDs all plateau at the same level for time-lags larger than 3\,s.
Only for shorter time-lags do the three regimes differ.}
\end{figure}

When a time-series of positions is examined, the first measure applied is frequently the mean-squared-displacent (MSD).  
This has to do with its ease of computation, the existence of exact analytical results for the expectation value, and comparative robustness of the measure to experimental measurement errors.
Also, there are no discretization effects to worry about---the MSD for finite sampling frequency is simply the MSD for continuous recording, evaluated at discrete time points.
More importantly, unlike the covariance and the power-spectra discussed below, the MSD is well-defined even if the process is not bounded.
In the present example, both the position and velocity processes are bounded, and we therefore have a simple relation between the MSD and the auto-covariance:
\beq \label{eq:MSDcovar}
   \mbox{MSD}(t) \equiv \la (x(t) -x(0))^2 \ra = 2\la x^2 \ra - 2 \la x(t) x(0) \ra \,
\eeq
where $\la x^2 \ra = \kB T /\kappa$, i.e., the MSD is a constant minus twice the auto-covariance of the position process (see Sec.~\ref{app:covariance}).
This is illustrated in Fig.~\ref{fig:AFM_MSD} where we show the MSD 
for the three time series plotted in Fig.~\ref{fig:AFM_positions}.

 \section{Covariance}
\label{app:covariance}
The stationary-state solution given in Eq.~(\ref{eq:OUtwosoln}) 
oscillates in resonating response to the thermal noise that drives it,
which is seen by its oscillating auto-covariance functions,
but its oscillations are randomly phase shifted by the same noise that drives oscillations,
which causes the exponential decay in auto-covariance functions, see below.
The result is that equal-time ensemble averages like $\langle x(t)^2\rangle$ are constant in time.
This is easily proven for quadratic expressions,
either directly from the solution given in Eq.~(\ref{eq:OUtwosoln}), 
or by differentiation w.r.t.\ time using It\={o}'s lemma:
\bea 
\lefteqn{\frac{\d}{\dt}
\left( \begin{array}{c}  \langle x^2 \rangle \\ \langle xv \rangle \\ \langle v^2 \rangle  \end{array} \right)
=}&& \\
&&\left( \begin{array}{ccc} 0      & 2             & 0   \\
                       -\omega_0^2 & \frac{-1}{\tau} & 1  \\
                                0 & -2\omega_0^2    & \frac{-2}{\tau} \end{array} \right)
\left( \begin{array}{c}  \langle x^2 \rangle \\ \langle xv \rangle \\ \langle v^2 \rangle  \end{array} \right)
+
\left( \begin{array}{c}  0 \\ 0 \\ \frac{2D}{\tau^2} \end{array} \right) \nonumber \e.
\label{eq:equaltimecovariance}
\eea
As is seen by inspection, this equation has the  time-independent solution
\beq
\left( \begin{array}{c}  \langle x^2 \rangle \\ \langle xv \rangle \\ \langle v^2 \rangle  \end{array} \right)
=
\left( \begin{array}{c}  \frac{\kT}{\kappa} \\ 0 \\ \frac{\kT}{m} \end{array} \right)
\enspace,
\eeq
in accord with the equipartition theorem.
This solution is the unique attractor for the system's dynamics:
Left to itself, any discrepancy from this time-independent solution
will decrease to zero exponentially fast in time---possibly
while oscillating harmonically with cyclic frequency $2\omega$---as is seen from the fact
that the $3\times3$-matrix in Eq.~(\ref{eq:equaltimecovariance})
has eigenvalues $-1/\tau$ and $-1/\tau \pm i 2\omega$, and determinant $-4 \omega_0^2/\tau$.

\begin{figure}
\includegraphics[width=\linewidth]{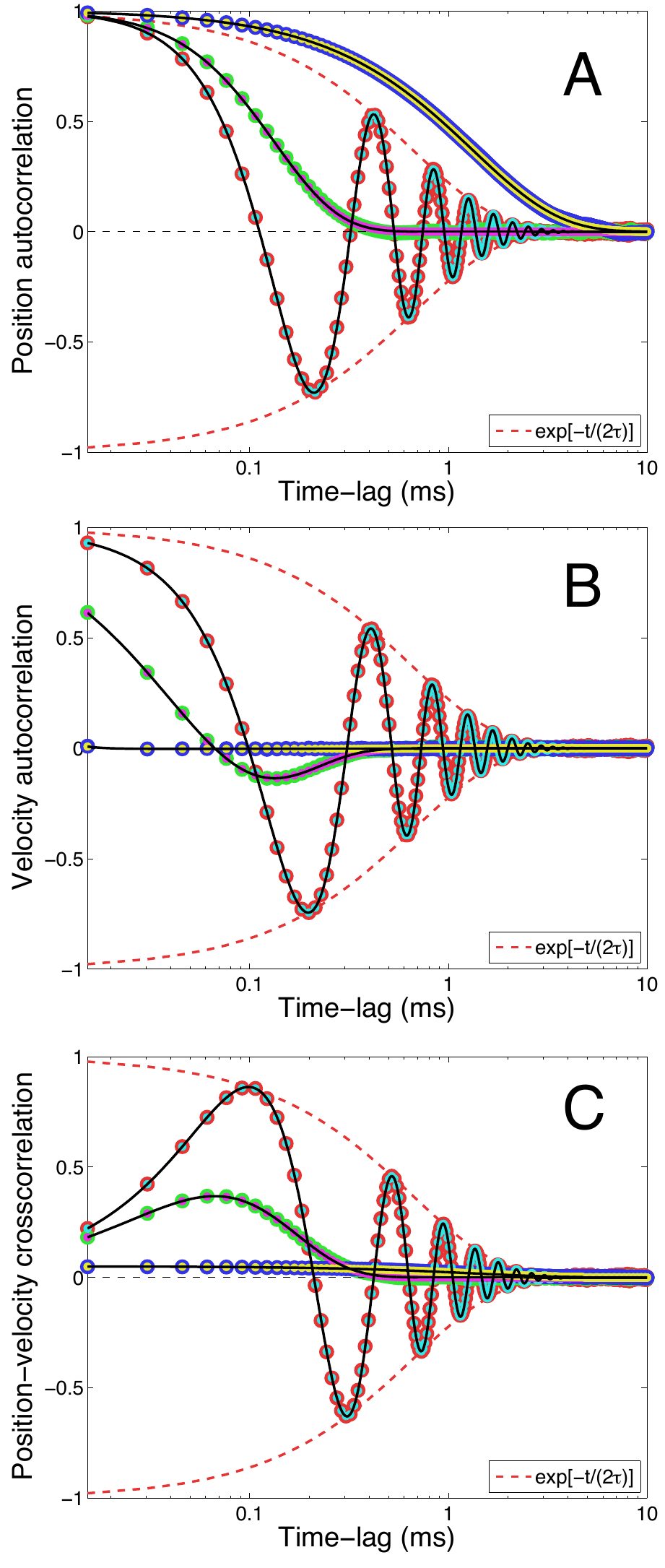}
\caption[]{\label{fig:AFM_corr}
Correlation functions, Eq.~(\ref{eq:covar}), for massive particles in a harmonic potential, driven by thermal noise.
Color-coding and simulation settings are the same as in Fig.~\ref{fig:AFM_MSD}:  Red, green, and blue show the result of a stochastic simulation; complementary colors show the covariance (not a fit), Eq.~(\ref{eq:covar}), normalized and evaluated at discrete time-points; thin black lines show the covariance for continuous time (not a fit); and dashed red lines show the enveloping exponential $\exp(-t/(2\tau)))$.
Panel A: Position auto-correlations, $\la x(t)x(0)\ra / \la x^2 \ra$.
Panel B: Velocity auto-correlations, $\la v(t)v(0)\ra /\la v^2 \ra$.
Panel C: Position-velocity cross-correlations, $\la x(t)v(0)\ra / \sqrt{ \la x^2\ra \la v^2 \ra }$.
The velocity-position cross-correlation, $\la v(t)x(0)\ra / \sqrt{ \la x^2\ra \la v^2 \ra }$, (not shown) has the opposite sign but is otherwise identical.}
\end{figure}

Similar reasoning (or simply multiplying Eq.~(\ref{eq:OUiter}) with $(x_j,v_j)$, then taking the expectation value, and applying the equipartition theorem) gives the covariances
\bea \label{eq:covar}
	\lefteqn{ \left( \begin{array}{cc}
	\la x(t) x(0) \ra & \la x(t) v(0) \ra \\
	\la v(t) x(0) \ra & \la v(t) v(0) \ra
	\end{array} \right) } && \\
	&&=
	\frac{D}{\tau}  e^{ -\frac{t}{2\tau} }
	\left( \begin{array}{cc}
	\frac{ \cos \omega t + \frac{\sin\omega t}{2\omega\tau} }{ \omega_0^2 }	& \frac{\sin \omega t}{\omega} \\
	\frac{-\sin\omega t}{\omega} & \cos \omega t - \frac{\sin\omega t}{2\omega\tau} \\
	\end{array} \right)	\nonumber \enspace.
\eea	
For $\omega$  real (underdamped system),
these covariances oscillate, we see, with amplitudes that decrease exponentially in time with characteristic time $2\tau$.
For $\omega$ imaginary (overdamped system), we rewrite Eq.~(\ref{eq:covar}) as
\bea
\lefteqn{ \left( \begin{array}{cc}
	\la x(t) x(0) \ra & \la x(t) v(0) \ra \\
	\la v(t) x(0) \ra & \la v(t) v(0) \ra
	\end{array} \right) } && \\
	&&=
	\frac{D}{2 |\omega| \tau}  \left\{
	\left( \begin{array}{cc}
	\frac{1 }{ \omega_0^2 \tau_+ }	& 1 \\
	-1 						& \frac{-1}{\tau_-} \\
	\end{array} \right) e^{-t/\tau_-}	
	+
	\left( \begin{array}{cc}
	\frac{-1 }{ \omega_0^2 \tau_- }	& -1 \\
	1 						& \frac{1}{\tau_+} \\
	\end{array} \right) e^{-t/\tau_+} \right\}
	\nonumber \enspace.
 \eea
from which we see that the system decreases as a double-exponential  with characteristic times $\tau_{\pm}=2\tau /(1 \pm 2\tau |\omega|)$.
At critical damping ($\omega=0$), the expressions for the covariances simplify to
\bea
\lefteqn{ \left( \begin{array}{cc}
	\la x(t) x(0) \ra & \la x(t) v(0) \ra \\
	\la v(t) x(0) \ra & \la v(t) v(0) \ra
	\end{array} \right) } && \\
	&&=
	\frac{D}{ \tau}  e^{ -\frac{t}{2\tau} }
	\left( \begin{array}{cc}
	\frac{1 + t/(2 \tau) }{ \omega_0^2  }	& t \\
	-t 						& 1-t/(2\tau) \\
	\end{array} \right)	\nonumber \enspace.
 \eea
Figure~\ref{fig:AFM_corr} illustrates these dynamics for the normalized covariances, aka the correlation functions.

\begin{figure}
\includegraphics[width=\linewidth]{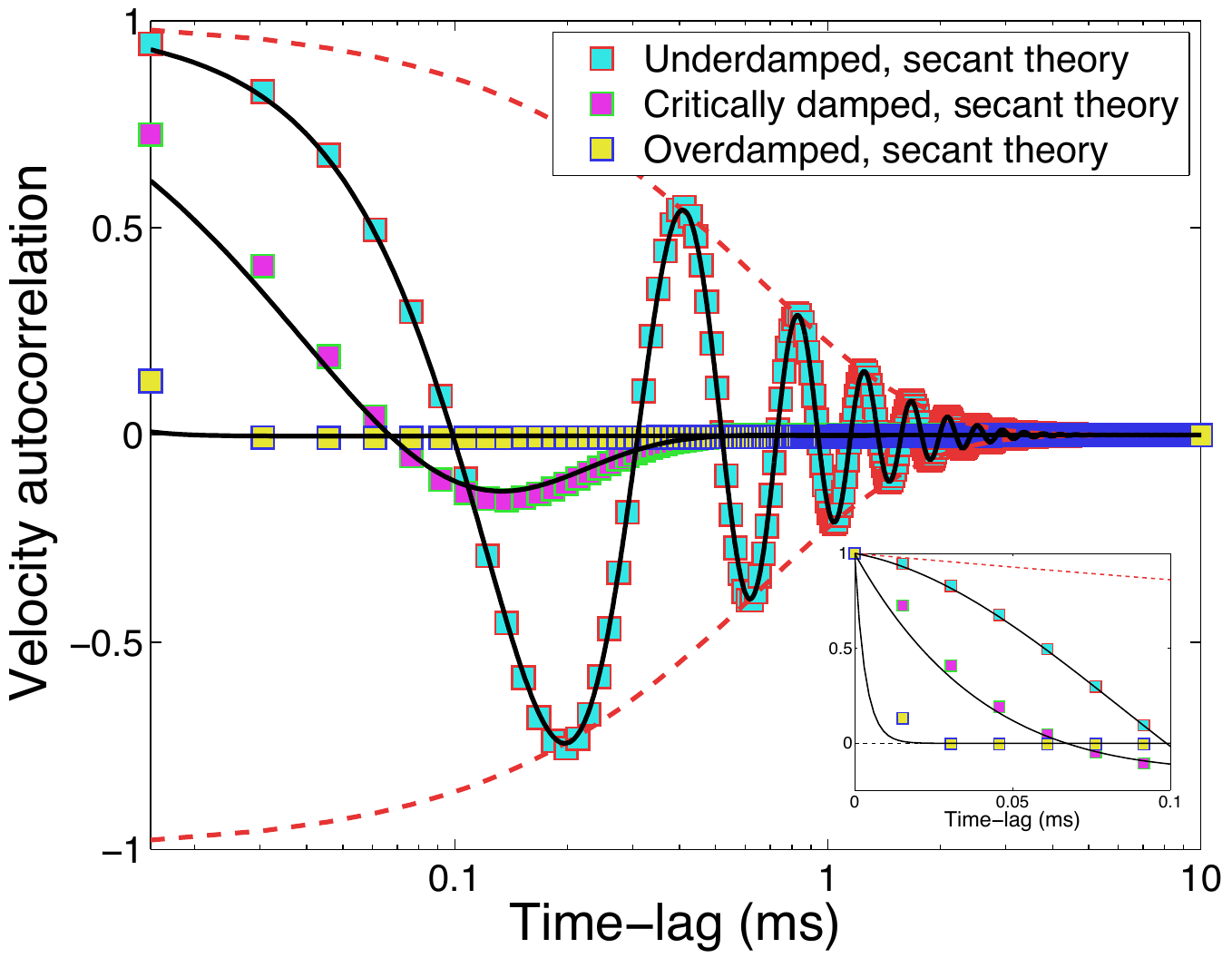}
\caption[]{\label{fig:AFM_secant_corr}
Secant-velocity autocorrelation functions for massive particles in a harmonic potential, driven by thermal noise.
Thin black lines, red dashed lines, and simulation settings are the same as in Fig.~\ref{fig:AFM_corr}B.
Additional symbols:  
Cyan-red, magenta-green, and yellow-blue squares show the autocorrelations for  $w_j = (x_j - x_{j-1} ) / \Dt$ calculated using $\la x(t)x(0)\ra$ from Eq.~(\ref{eq:covar}).
Insert shows the first 0.1\,ms on a linear time-scale, revealing the very fast decay of the overdamped oscillator and how the secant-velocity autocorrelations overestimate the instantaneous-velocity autocorrelation decay-times.  
As expected, the secant-velocity does a better job estimating the instantaneous velocity for the less damped system because of  its smoother particle-trajectory.
}\end{figure}

As is the case for the MSD, there are no discretization effects to worry about---the covariance for finite sampling frequency is simply the covariance for continuous recording, evaluated at discrete time points.
However, the given velocity correlations, and thus also the position-velocity correlations, are only correct for the actual, instantaneous velocities:
If the instantaneous velocity is not directly measured, but instead estimated from measured positions as a ``secant-velocity'' (see Eq.~(\ref{eq:secant_vel}) and Sec.~\ref{sec:secantvelocities}), then the corresponding secant-velocity correlation function should be calculated from the position covariance function given in Eq.~(\ref{eq:covar}).  Figure~\ref{fig:AFM_secant_corr} shows how such secant-velocity correlations  deviate from the instantaneous-velocity correlations, mainly for short time-lags.

\section{Power spectra}
The power spectral density (PSD) of a variable $x(t)$ or $v(t)$ is defined as the expectation value of the squared modulus of its Fourier transform, with a normalization that differs between authors. 
We choose one in which the power spectral density remains constant for increasing measurement time, $\Tmsr$ 
\beq \label{eq:AFMPSDx}
	P^{(x)}(f_k) \equiv \la | \tilde{x}_k |^2 \ra / \Tmsr = \frac{ D/(2 \pi^2) }{ (2 \pi \tau )^2 (f_k^2 - f_0^2)^2 + f_k^2 }
\eeq
and
\bea  \label{eq:AFMPSDv}
	P^{(v)}(f_k) &\equiv& \la |\tilde{v}_k |^2 \ra / \Tmsr = (2\pi f_k)^2 P^{(x)}(f_k) \nonumber \\
	&=&	\frac{ 2D }{ (2\pi\tau)^2 (f_k - f^2_0/f)^2 + 1 } \e,
\eea
where $2\pi f_0 = \omega_0$ is the resonance frequency of the system.
Here, we used $\tilde{v}_k = i2\pi \tilde{x}_k$, 
which is an approximation that ignores contributions from the finite measurement-time.  

The finite-time Fourier transform (FTFT) used above is defined as
\beq \label{eq:finiteFT}
	\tilde{x}_k = \int_{0}^{\Tmsr} \! \dt \,  e^{i2\pi f_k t} \, x(t), \,\, f_k\equiv k/\Tmsr,\,\, k\,\, \mbox{integer} \e.
\eeq
In a similar manner we can calculate the finite-sampling-frequency PSDs  by discrete Fourier transformation (DFT) of Eq.~(\ref{eq:OUiter})
\beq
\label{eq:OUFT}
 	e^{-i \pi f_k / \fNyq}\left( \begin{array}{c}  \hat{x}_{k} \\ \hat{v}_{k}  \end{array} \right) =
 	e^{- \mathbf{M}\Delta t}   \left( \begin{array}{c}  \hat{x}_{k} \\ \hat{v}_{k}  \end{array} \right)
	+  \left( \begin{array}{c}  \Delta \hat{x}_k \\ \Delta \hat{v}_k    \end{array} \right)
	\enspace,
\eeq
where
\beq \label{eq:DFT}
	\hat{z}_k 	= \Dt \sum_{j=1}^N e^{i2\pi j k/N} z_j 
	= \frac{2}{\fNyq}  \sum_{j=1}^N e^{i\pi j f_k/\fNyq} z_j \e,
\eeq
for $z=x,v,\Dx,\Dv$; $x_{N+1} = x_1$ and $v_{N+1} = v_1$; 
\bea
	f_k 		&=& k \Delta f = k / \Tmsr \\
	k 		&=& 0, 1, 2, \ldots  ,N-1 \\
	\Tmsr	&=& N \, \Dt = N / \fsample \\
	\fNyq		&=& \frac{N}{2} \Delta f = \frac{\fsample}{2} \e,
\eea
and, 
\bea
	\la \Delta \hat{x}_k^* \Delta \hat{x}_{k'} \ra
	&=& \sigma_{xx}^2 \, \la \hat{\xi}_k^*  \hat{\xi}_{k'} \ra
	= \sigma^2_{xx} \, \Dt \,\Tmsr \,\delta_{k,k'} \\
	\la \Delta \hat{v}_k^* \Delta \hat{v}_{k'} \ra
	&=& \sigma^2_{vv} \, \Dt \,\Tmsr \,\delta_{k,k'} \\
	\la \Delta \hat{x}_k^* \Delta \hat{v}_{k'} \ra
	&=& \sigma^2_{xv} \, \Dt \,\Tmsr \,\delta_{k,k'} \e.
\eea
After isolating $(\hat{x}_k, \hat{v}_k)$ in Eq.~(\ref{eq:OUFT}), we find
\beq
	\left( \begin{array}{cc}
	\la |\hat{x}_k|^2 \ra  & \la \hat{x}_k \hat{v}_k^* \ra\\
	\la \hat{x}_k^* \hat{v}_k \ra & \la | \hat{v}_k|^2 \ra
	\end{array} \right) 
	= \mathbf{A}_k^{-1} 
	\left( \begin{array}{cc}
	\sigma_x^2 & \sigma_{xv}^2 \\
	\sigma_{xv}^2 & \sigma_v^2 \\
	\end{array} \right) 
	 (\mathbf{A}_k^\dagger)^{-1} \,\Tmsr \Dt
\eeq
where the $2\times 2$ matrix and its inverse are
\beq
	\mathbf{A}_k = \alpha_k \mathbf{I} + \beta  \mathbf{J} \, ; \,\,\,
	\mathbf{A}_k^{-1} = \frac{ \alpha_k \mathbf{I} - \beta   \mathbf{J}  }{ \alpha_k^2 + \beta^2 }
\eeq
with
\bea
	\alpha_k 	&=& e^{-i\pi f_k / \fNyq} - e^{ -\frac{\Dt}{2\tau} } \cos(\omega \Dt) \\
	\beta		&=& -e^{ -\frac{\Dt}{2\tau} } \sin(\omega \Dt) \enspace,
\eea
complex scalars ($\beta/\omega$ is real), Hermitian conjugation is denoted by $\dagger$, and complex conjugation by $*$.

For the discrete positional power spectrum we thus get (see also \cite[Appendix A]{Norrelykke2010})
\bea  \label{eq:PSDx}
	P_k^{(x)} &\equiv&  \la | \hat{x}_k |^2 \ra/\Tmsr \\
	&=&\frac{  | \alpha_k - \frac{\beta}{2\omega\tau} |^2 \sigma^2_{xx}
	- 2 \frac{\beta}{\omega} \, \mbox{Re} \left\{ \alpha_k - \frac{\beta}{2\omega\tau}  \right\} \sigma^2_{xv}
	+ \frac{\beta^2}{\omega^2}  \sigma^2_{vv}
	}{ | \alpha_k^2 + \beta^2 |^{2} \, \fsample}  \nonumber
\eea
whereas the discrete velocity power spectrum is
\bea \label{eq:PSDv}
	&&P_k^{(v)} \equiv \la | \hat{v}_k |^2 \ra/\Tmsr \\
	&=& \frac{  | \alpha_k + \frac{\beta}{2\omega\tau} |^2 \sigma^2_{vv}
	+ 2 \omega_0^2 \frac{\beta}{\omega} \, \mbox{Re}\left\{ \alpha_k + \frac{\beta}{2\omega\tau}  \right\} \sigma^2_{xv}
	+ \omega_0^4  \frac{\beta ^2}{\omega^2}  \sigma^2_{xx}
	}{ | \alpha_k^2 + \beta^2 |^{2} \, \fsample } \nonumber \enspace .
\eea
In case of critical damping ($\omega = 0$) we simply insert $\beta/\omega = - e^{-\frac{\Dt}{2\tau}} \Dt$ in Eqs.~(\ref{eq:PSDx}) and (\ref{eq:PSDv}), and use Eqs.~(\ref{eq:sxxcrit},\ref{eq:svvcrit},\ref{eq:sxvcrit}) for the variance-covariances.

\begin{figure}
\includegraphics[width=\linewidth]{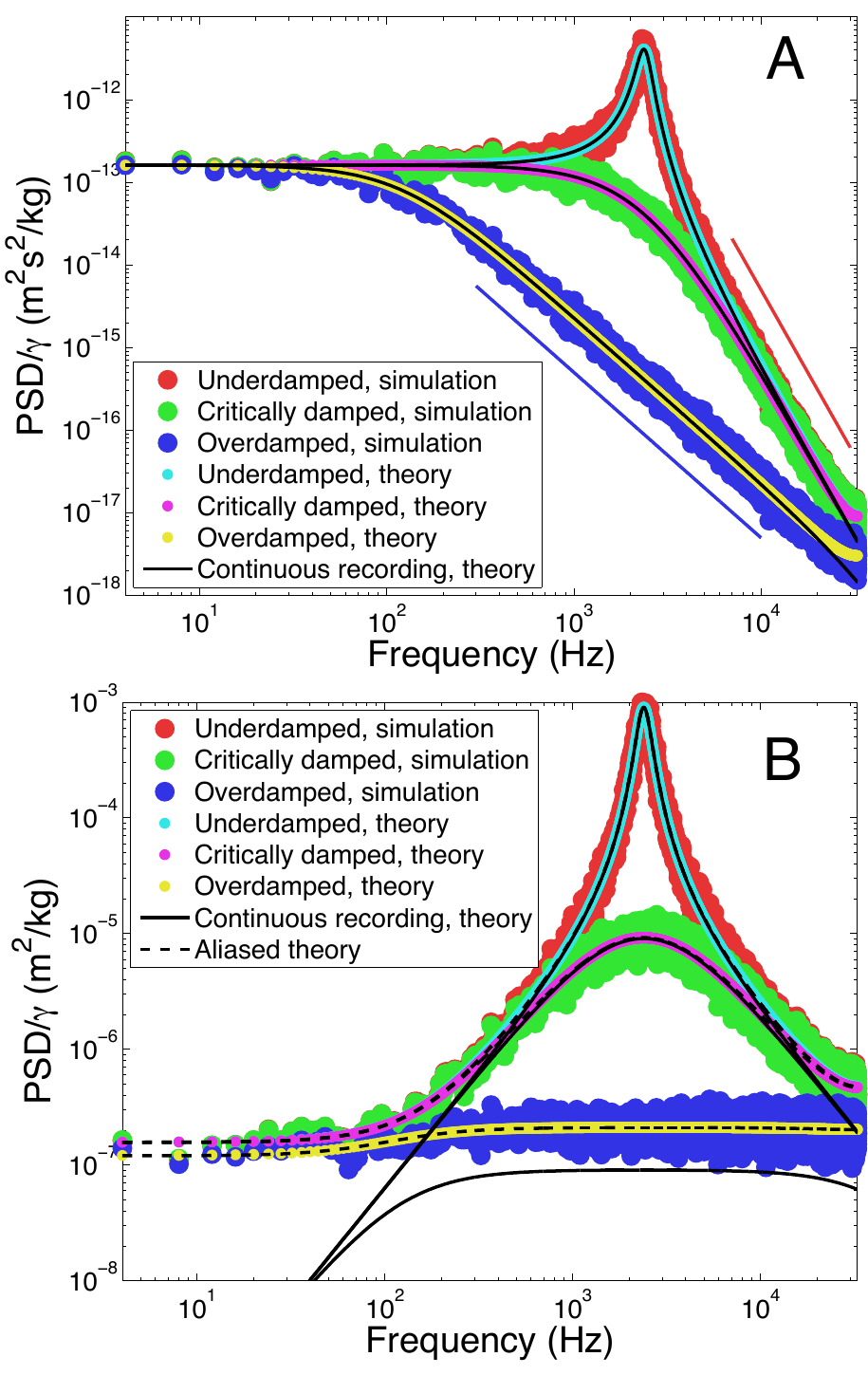}
\caption[]{\label{fig:AFM_PSDs}
Power spectra for positions and velocities of harmonically trapped massive particles in thermal bath, normalized by $\gamma$.   
The color-scheme is the same as in Figs.~\ref{fig:AFM_MSD} and \ref{fig:AFM_corr}: Synthetic data is shown in red, green, and blue;
the aliased theory, Eqs.~(\ref{eq:PSDx}) and (\ref{eq:PSDv}), is shown in complementary colors; and the non-aliased theory, Eqs.~(\ref{eq:AFMPSDx}) and (\ref{eq:AFMPSDv}), as thin black lines. 
Simulation parameters are as in Fig.~\ref{fig:AFM_positions} (giving $\omega_0 = 15$\,kHz) with $\nwin = 32$ Hann windows applied to the  $\Tmsr = 8$\,s long time-series.
Panel A: Power spectra for positions, $f^{-2}$ and $f^{-4}$ behaviors are illustrated by the thin blue and red lines, respectively.
Notice the disagreement between the non-aliased (continuous recording) theory and data at high frequencies. 
Panel B: Power spectra for velocities.  Dashed black lines show aliased versions of Eq.~(\ref{eq:AFMPSDv}), i.e., $P^{(v,\rm aliased)}(f) = \sum_{n=-\infty}^{\infty}P^{(v)}(f + n \fsample)$, \cite{BergSorensen2004}, here truncated to 201 terms. The non-aliased theory severely underestimates the power at high and low frequencies, and completely misses the data in the overdamped case.}
\end{figure}

Figure~\ref{fig:AFM_PSDs} shows the power spectra that result from numerical simulations of AFM cantilever positions and velocities using Eq.~(\ref{eq:OUiter}) for three different drag coefficients, as well as the corresponding analytical expressions as given in Eqs.~(\ref{eq:PSDx},\ref{eq:PSDv}).

An alternative route to these discrete PSDs is to take the discrete Fourier transform of the covariance function from the previous section:  Both the position and the velocity processes are stationary (the joint probability density function  for each is independent of time), so the Wiener-Khinchin theorem applies and the Fourier transform of the correlation function is equal to the PSD\@. 
For vanishing spring constant, $\kappa$, the position process is unbounded (not stationary) and the covariance, as well as the PSD, are consequently ill-defined. This limit is treated in Sec.~\ref{sec:OU}.

Note, the PSD for discretely measured (instantaneous) velocities, Eq.~(\ref{eq:PSDv}), should not be confused with the discrete PSD for secant-velocities, Eq.~(\ref{eq:PSDw}). The latter is the correct expression to use in experiments where the velocities are estimated from the measured positions.

\section{Vanishing $\Dt$: The limit of continuous recording}
As $\Dt \rightarrow 0$  we find to first order in $\Dt$ that the covariances in Eqs.~(\ref{eq:sxx}--\ref{eq:sxvcrit}) reduce to 
$\sigma_{xx}^2 = 0$, 
$\sigma_{vv}^2 =  2D\Dt / \tau^2$, and $\sigma_{xv}^2 = 0$.
That is, $\Dx_j = 0$ and $\Dv_j = \sigma_{vv} \zeta_j = \sqrt{ 2 D \Dt } /\tau \,\zeta_j$.
In other words, the velocity process is seen to be driving the position process.

In this limit the positional PSD takes on the familiar form given in Eq.~(\ref{eq:AFMPSDx}), and the velocity PSD is given in Eq.~(\ref{eq:AFMPSDv}).
Comparing the continuous recording PSDs Eqs.~(\ref{eq:AFMPSDx},\ref{eq:AFMPSDv}) with their discrete sampling analogues Eqs.~(\ref{eq:PSDx},\ref{eq:PSDv}) we see in Fig.~\ref{fig:AFM_PSDs}A that they differ substantially at high frequencies, and in Fig.~\ref{fig:AFM_PSDs}B that they differ everywhere but near $f_0$ in the underdamped case.
Thus, one should not attempt to fit a theoretical PSD derived for ``continuously recorded'' trajectories
to an experimental PSD, which necessarily is obtained with time-lapse recording.  
More specifically, one should only attempt to fit it to the low-frequency part of the experimental spectrum for positions, or account for ``aliasing" before fitting, as described e.g.\ in \cite[Appendix H]{BergSorensen2004}, which turns Eqs.~(\ref{eq:AFMPSDx}) and (\ref{eq:AFMPSDv}) into Eqs.~(\ref{eq:PSDx}), respectively (\ref{eq:PSDv}).
 
The position-covariance and the MSD do not suffer from this dichotomy.
None of them change their shape or form when switching between discrete and continuous time:
The discrete versions are equal to the continuous versions, evaluated at discrete times:
\beq
	\la x_j x_{j+\ell} \ra = \la x(t_j) x(t_{j+\ell}) \ra = \la x(0)  x(t_{\ell})  \ra
\eeq
and
\beq
	\la (x_j - x_{j+\ell} )^2 \ra = \la (x(t_j) - x(t_{j+\ell}) )^2 \ra = \la (x(0) - x(t_{\ell}) )^2 \ra \e,
\eeq
where, in the last step, we used that the process is stationary and the measures therefore invariant to time-translations.
That is, these measures are unaffected by the unavoidable discreteness of real-world data.

The velocity's covariance is also unaffected by time-lapse recording, if we can measure the instantaneous velocity, i.e. the velocity vector that is tangential to the trajectory of the position.  If we cannot and time-lapse record only positions, we have a different situation, which is treated below.

\section{Various physical limits:  A reference-set of formulas}
Three physical limiting cases are of particular interest:
(i) Vanishing spring constant, $\kappa=0$, which is the Ornstein-Uhlenbeck theory for Brownian motion for a free particle with inertia;  
(ii) vanishing mass, $m=0$, which is Einstein's theory for Brownian motion of a particle trapped by a Hookean force, and a popular minimalist model for the Brownian motion of a microsphere in an optical trap; 
and (iii) vanishing mass and spring constant, which is Einstein's theory for Brownian motion of a free particle.

\subsection{Vanishing spring constant:\\ The Ornstein-Uhlenbeck process}
\label{sec:OU}
When there is no Hookean restoring force $\kappa = 0$, Eq.~(\ref{eq:Newton}) describes the
free diffusion of a massive particle  according to Ornstein and Uhlenbeck \cite{Uhlenbeck1930}
\beq  \label{eq:OU}
	m\dot{v}(t) + \gamma v(t) = \Ftherm(t) \e.
\eeq
This velocity-process is known as the Ornstein-Uhlenbeck (OU) process \cite{Uhlenbeck1930}.
When modeling other dynamical systems, such as migrating cells, $m$ does not refer to the physical mass of the cell but rather its inertia to velocity-changes, or persistence of motion; likewise $\gamma$ is not the friction between the cell and substrate but describes the rate of memory-loss for the velocity process.
The structure of Eqs.~(\ref{eq:OUiter},\ref{eq:expM},\ref{eq:IJ},\ref{eq:Dx2},\ref{eq:Dv2}) remain the same, the only change is $\omega_0 = 0$ in Eq.~(\ref{eq:IJ}), hence $\omega=i/(2\tau)$, and the covariances, Eqs.~(\ref{eq:sxx},\ref{eq:svv},\ref{eq:sxv}), consequently reduce to
\bea
	\sigma^2_{xx} 	&=& D \tau \left(   2\Dt/\tau - 4(1 - a) + (1- a^2)  \right) \label{eq:sxxOU}\\
	\sigma^2_{vv} 	&=& D/\tau \left( 1 - a^2 \right) \label{eq:svvOU} \\
	\sigma^2_{xv}	&=& D \left(  1  - a \right)^2 \label{eq:sxvOU} \enspace,
\eea
where
\beq
	a = e^{-\Dt/\tau} , \,\,\,  \tau =m/\gamma
 	\enspace.
\eeq
For the velocity and position processes we thus find
\bea
	v_{j+1} 	&=& a v_j + \Dv_j \label{eq:OUvdisc}\\
	x_{j+1} 	&=& x_j + \tau (1-a) v_j + \Dx_j \label{eq:OUxdisc} \enspace.
\eea
That is, the velocity process has reduced to a stable autoregressive model of order one, AR(1).
Its time integral, the position process, is unbounded---the particle is diffusing freely and without limits.
Exact numerical update formulas for $v_j$ and $x_j$ have previously been given in \cite{Gillespie1996}, here we re-derived them for consistency of notation.

The discrete-time positional PSD, $P_k^{(x)}$, is obtainable from Eq.~(\ref{eq:PSDx}) with $\omega_0 = 0$ and the $\sigma$s given in Eqs.~(\ref{eq:sxxOU},\ref{eq:svvOU},\ref{eq:sxvOU}); the expression does not simplify significantly compared to Eq.~(\ref{eq:PSDx}).
The discrete-time velocity PSD, $P_k^{(v)}$, is much simpler and can be derived directly from the discrete-time velocity process, Eq.~(\ref{eq:OUvdisc}); or by taking the $\kappa=0$ limit of Eq.~(\ref{eq:PSDv}).
From these discrete-time PSDs the continuous-recording expressions, $P^{(x)}(f)$ and $P^{(v)}(f)$, can be obtained by expanding to leading order in $\Dt/\tau$ and $k/N=f_k/\fsample$; or they can be derived directly from the continuous-time equation of motion Eq.~(\ref{eq:OU}) by Fourier transformation and, for the positional PSD, remembering that $\dot{x}=v$.
The continuous-recording positional PSD and the two velocity PSDs then read
\bea
	P^{(x)}(f_k) &=&  \frac{ D / (2\pi^2) }{  (2 \pi \tau)^2 f_k^4  + f_k^2}  \label{eq:PSDxOU} \\
	P_k^{(v)} &=&  \frac{\sigma_{vv}^2 \, \, \Dt}{1+ a^2 - 2 a \cos(2\pi k/N)}  \label{eq:PSDvOUalias}\\
	P^{(v)}(f_k)&=&  \frac{ 2D}{ 1 + (2\pi f_k \tau)^2 } \label{eq:PSDvOU} \enspace.
\eea
Note, that the positional PSD diverges in $f_k=0$ due to the process being unbounded.
This implies that we cannot use the Wiener-Khinchin theorem to obtain the covariance by Fourier transforming the PSDs, and vice versa.
The velocity process is bounded, however, and the velocity correlation functions are straightforward to calculate, so we simply list the results for the discrete and the continuous case
\bea
	\la v_i v_j \ra &=&  \frac{ D }{ \tau } \,\, e^{ - | i-j | \Dt / \tau} \label{eq:cvvOU} \\
	\la v(t) v(t') \ra &=& \frac{ D }{ \tau } \,\, e^{-|t-t'|/\tau} \enspace.
\eea
As already mentioned, the time integral of the OU-process is one of the instances where the mean-squared-displacement provides a useful measure for the position process, while the positional covariance function is ill-defined.
The mean-squared-displacement of the time integral of the OU-process is
\beq
	\la (x(t) -x(0))^2 \ra = 2D \tau \left( t/\tau + e^{-t / \tau} -1 \right) \e.
\eeq
For $t \ll \tau$ this MSD increases as $t^2$ (ballistically) and for $t \gg \tau$ as $t$ (diffusively), with an exponential cross-over between the two regimes with characteristic time $\tau$.

\subsubsection{Secant-velocities}
\label{sec:secantvelocities}
In experiments, the velocity is typically not measured directly, but approximated from the measured positions as a ``secant-velocity'':
\beq \label{eq:secant_vel}
	w_j = (x_j - x_{j-1} ) / \Dt \,
\eeq
with discrete power spectral density
\bea \label{eq:PSDw}
	P^{(w)}_k &\equiv& \la | \hat{w}_k| ^2 \ra / \Tmsr \nonumber \\
	&=& \frac{2\left( 1 - \cos ( \pi f_k / \fNyq ) \right)}{(\Dt)^2}  \, P_k^{(x)} \e.
\eea
This is a general result that follows from the definition of $w_j$ and is independent of the dynamic model.

\begin{figure}
\includegraphics[width=\linewidth]{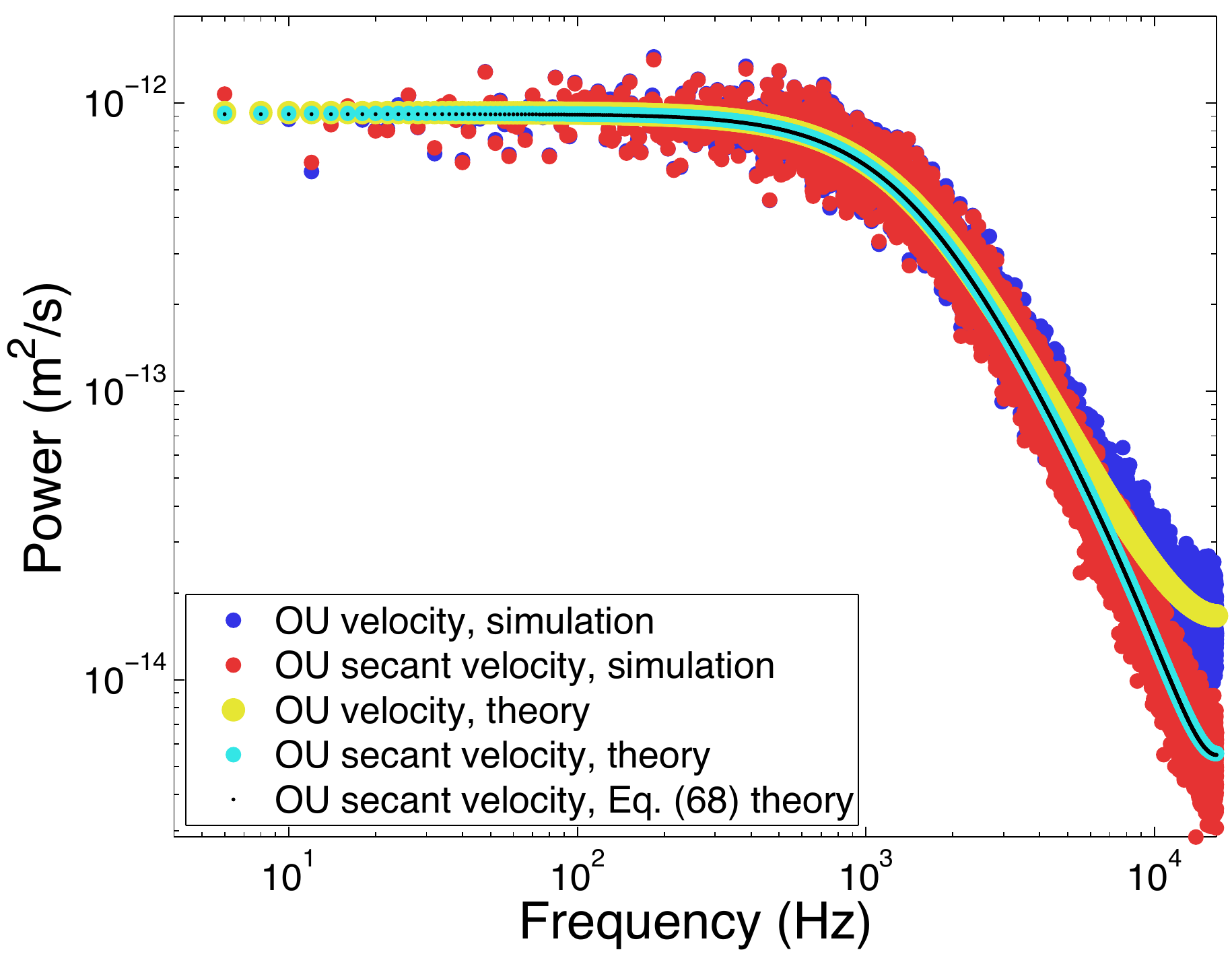}
\caption[]{\label{fig:PSDw}
Power spectra of velocities and secant-velocities for the OU process.
Blue:  Simulated velocities, Eq.~(\ref{eq:OUvdisc}). 
Red: Secant-velocities, Eq.~(\ref{eq:secant_vel}), calculated from simulated positions, Eq.~(\ref{eq:OUxdisc}). 
Yellow: Aliased velocity theory, Eq.~(\ref{eq:PSDvOUalias}). Cyan:  Aliased secant-velocity theory, Eq.~(\ref{eq:PSDw}).
Black line: Approximated aliased secant-velocity theory, Eq.~(\ref{eq:PSDwapprox}).
Simulation settings are given the caption of Fig.~\ref{fig:OT_PSDs}.
Notice the discrepancy between the true velocity and the estimated (secant) velocity at high frequencies.
}
\end{figure}

For the OU-process, we can relate this PSD to the PSD for the continuous velocity that it approximates, if we rewrite it as
\bea
	P^{(w)}_k  &=& P^{(v)}_k \, (\tau/\Dt)^2 (1-a)^2 + \sigma_{xx}^2 / \Dt \nonumber \\
	&+&\tau(1-a)\left( \la\hat{v}_k^* \, \Delta \hat{x}_k \ra + c.c. \right) \e.
\eea
Here the first term is proportional to $P^{(v)}$, the second term is a constant, and the last term is frequency dependent with
$c.c.$ the complex conjugate of the other term in the bracket.  However, it is not easy to see from this expression how much $P^{(w)}$ deviates from $P^{(v)}$.
We can get a good idea about the shape of the PSD from the auto-correlation function
\bea
	\la w_j^2 \ra &=& 2 \la v_j^2 \ra  \left( \frac{ \tau }{ \Dt } \right)^2 \left( \Dt/\tau -1 + a \right) \label{eq:wwl}\\
	& \approx & \la v_j^2 \ra \left( 1 - \frac{1}{3} \Dt/\tau \right) \e,
\eea
and for $\ell > 0$
\bea
	\la w_j w_{j+\ell} \ra &=& 2 \la v_j v_{j+\ell} \ra  \left( \frac{ \tau }{ \Dt } \right)^2 \left( \cosh(\Dt/\tau) - 1 \right) \label{eq:wwl0}\\
	& \approx & \la v_j v_{j+\ell} \ra \left( 1 + \frac{1}{12} (\Dt/\tau)^2  \right) \e,
\eea
where we used Eqs.~(\ref{eq:Dx2},\ref{eq:Dv2},\ref{eq:sxvOU},\ref{eq:OUxdisc},\ref{eq:cvvOU}) to derive Eqs.~(\ref{eq:wwl},\ref{eq:wwl0})---the latter two expressions are proportional to the velocity autocorrelations they approximate, they only differ by multiplicative factors that are independent of the time-lag for $\ell>0$. 
To first order in $\Dt/\tau$ we thus have, for $\ell \ge 0$, 
\beq
	\la w_j w_{j+\ell} \ra = \la v_j v_{j+\ell} \ra \left( 1 -  \frac{1}{3} \frac{\Dt}{\tau} \delta_{0,\ell} \right) \e.
\eeq 
We now apply the Wiener-Khinchin theorem to get the PSD as the Fourier transform of this approximated auto-correlation function and find
\beq \label{eq:PSDwapprox}
	P^{(w)}_k \approx P^{(v)}_k - \frac{1}{3} \left( \frac{ \Dt }{ \tau } \right)^2  D \e.
\eeq
This approximation to the secant-velocity PSD, as well as the exact expression Eq.~(\ref{eq:PSDw}), are both shown in Fig.~\ref{fig:PSDw}, together with numerical simulation results and the PSD of the continuous velocity Eq.~(\ref{eq:PSDvOUalias}).  
Notice, that even with $\Dt / \tau = 0.27 < 1$ ($\Dt = 1/\fsample$ and $\tau = m/\gamma$), the relative difference between $P^{(v)}_k$ and $P^{(w)}_k$ is substantial for higher frequencies.
Thus, proper care should be taken if empirical testing of a model includes the fitting of a theoretical velocity PSD to an experimental secant-velocity PSD, even when aliasing is properly accounted for.
But, we now also know how to do this correctly:  Either add $1/3 (\Dt/\tau)^2 D$ to the experimental secant-velocity PSD before fitting with $P^{(v)}_k$, or fit directly using the full theoretical expression for $P^{(w)}_k$ given in Eq.~(\ref{eq:PSDw}).
In cases where this corrective constant is much smaller than the velocity PSD \emph{for all frequencies fitted}, it can obviously be ignored.

We observe that the secant velocity is the time-average of the real velocity in the time interval spanned by the secant.  
Time-averaging is low-pass filtering, as is well known \cite[Chapter 13]{Press1992} and demonstrated in Fig.~\ref{fig:PSDw}.

\subsection{Vanishing mass:\\ The optical trap limit}
\label{sec:OT}
The \emph{position} process in the limit of vanishing mass is mathematically identical to the OU \emph{velocity} process in the limit of vanishing trapping force treated above. 
Equation~(\ref{eq:Newton}) reduces to
\beq  \label{eq:OT}
	\gamma \dot{x}(t) + \kappa x(t) = \Ftherm(t)
\eeq
with solution
\beq
	x(t) = \frac{1}{\gamma} \int_{-\infty}^t \dt' \, e^{-2\pi f_c (t-t')} \, \Ftherm(t') \e,
\eeq
where the corner frequency $\fc \equiv \kappa / (2 \pi \gamma)$ is the frequency where $P^{(x)}(f) = P^{(x)}(0)/2$, see below.
Recycling the results from Section~\ref{sec:OU} we can directly write down the autocovariance
\beq \label{eq:OTautocov}
	\LA x(t) x(t') \RA = \la x^2 \ra \, e^{-2 \pi \fc |t - t'|}
\eeq
with  $\la x^2 \ra = \kT / \kappa$ and the discrete update rules
\bea
	x_{j+1}	&=& c \, x_j + \Delta x_j  \label{eq:OTxi}\\
	c 		&=& \exp( - \kappa \Dt / \gamma ) \\
	\Dx_j	&=& \sqrt{ \frac{(1-c^2) D \gamma }{\kappa} } \, \xi_j \e,
\eea
where $\xi$ are uncorrelated random numbers of unit variance, zero mean, and Guassian distribution.
Likewise, the position PSDs for discrete sampling, respectively, continuous recording are found to be
\bea
 	P_k^{(x)} 	&=& \frac{ D \, (1-c^2)\,\Dt \gamma/\kappa  }{ 1 + c^2 - 2c\cos(2 \pi k / N) } \\
	P^{(x)}(f_k) 	&=& \frac{ D/(2 \pi^2) }{\fc^2 + f_k^2}  \label{eq:OTxPSD}\e.
\eea

Finally, the mean-squared-displacement is 
\beq
	\la (x(t) -x(0))^2 \ra = 2D\gamma/\kappa \left( 1 - e^{-t\gamma/\kappa} \right) \e,
\eeq
which shows an exponential cross-over from linear dependence on $t$ to the  constant value $2D\gamma/\kappa$ as $t \gg \kappa / \gamma$, see Fig.~\ref{fig:AFM_MSD}.

\subsection{Vanishing mass and spring constant:\\ Einstein's Brownian motion}
\label{sec:BM}
When both $m=0$ and $\kappa=0$, Eq.~(\ref{eq:Newton}) reduces to
\beq  \label{eq:Langevin}
	 \gamma \dot{x}(t) = \Ftherm(t)
\eeq
so that
\beq
	x_{j+1} = x_j + \sqrt{2D\,\Dt } \,\, \xi_j \enspace,
\eeq
where
\beq
	\xi_j \equiv  \frac{1}{\sqrt{\Dt}} \int_{t_j}^{t_{j+1}} \!\!\! \dt\, \eta(t) \e,
\eeq
hence
\beq
	 \la \xi_j \ra =  0,\,\, \la \xi_i \xi_j \ra =  \delta_{i,j} \e,
\eeq
and the velocity and position PSDs for discrete and continuous sampling take on the simple forms:
\bea
	P_k^{(x)} &=&  
	\frac{ D/\fsample^2}{ 2\sin^2(\pi f_k /\fsample)} \\
	P^{(x)}(f_k) &=&  \frac{D}{ 2 \pi^2 f_k^2}\\
	P^{(v)}_k &=& P^{(v)}(f_k) = 2D \e.
\eea
As was the case for the time integral of the OU process, the position PSDs are singular in $f_k=0$ because the process is unbounded.  That is, the meaningful measure to study is not the covariance, but rather the mean-squared-displacement 
\beq
	\la (x(t) -x(0))^2 \ra = 2Dt \e,
\eeq
which is one of the few well-defined statistics for Einstein's theory for Brownian motion:
Because the trajectory of positions is a fractal, attempt at estimating the average speed of Brownian motion from the displacement of position occurring in a given time-interval  will depend on the duration, $\Dt$, of this interval
as $1/\sqrt{\Delta t}$, hence diverge when accuracy is sought improved by reducing $\Dt$.
This was not appreciated before Einstein's 1905 paper on the subject.

\begin{figure}
\includegraphics[width=\linewidth]{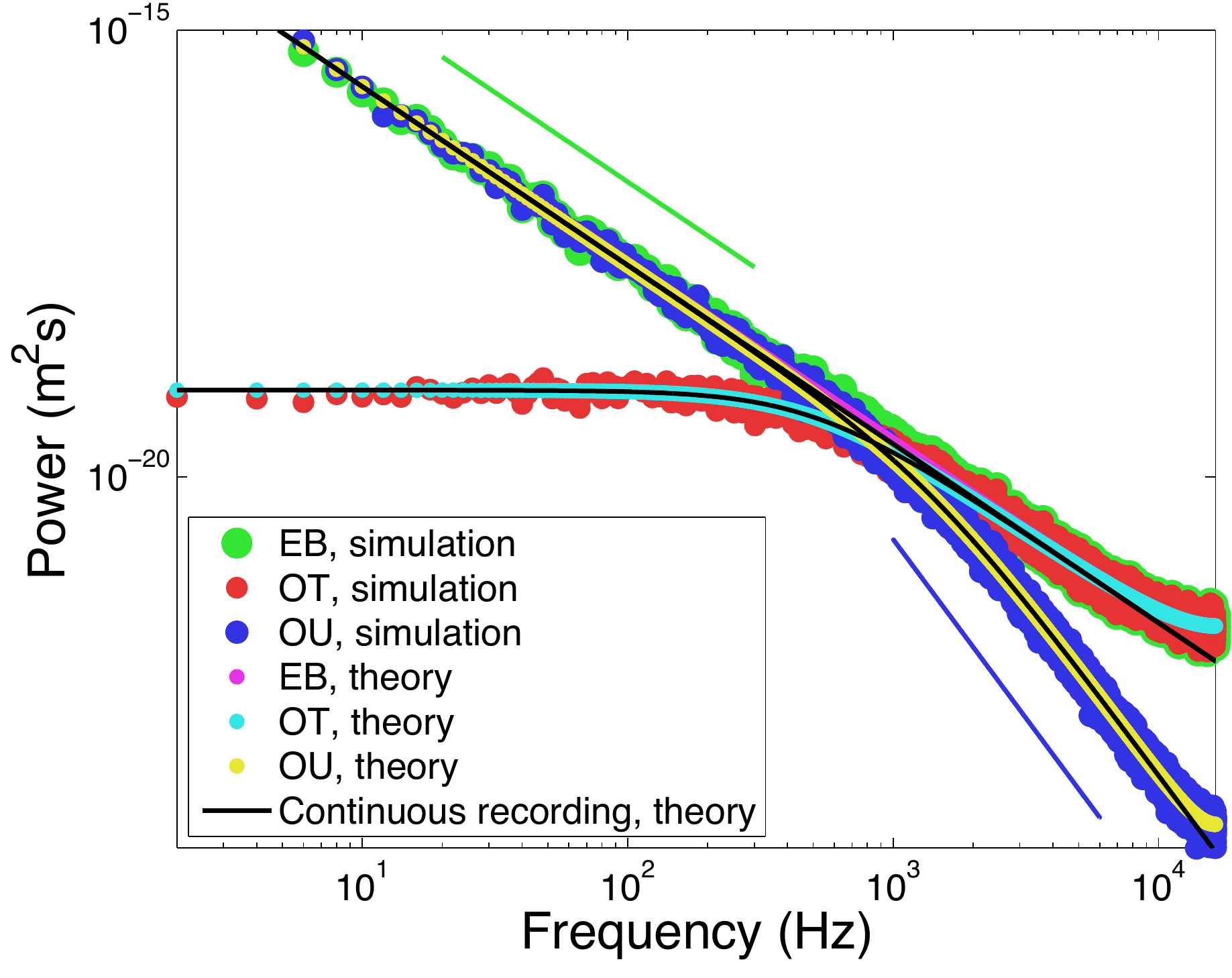}
\caption[]{\label{fig:OT_PSDs}
Power spectra of positions.
Green points: Freely diffusing massless particle (Einstein's Brownian motion); 
Red points: Trapped massless particle (OT limit, or OU velocity process);  
and Blue points: Freely diffusing massive particles (time integral of OU process).
Complementary colors show the finite sampling frequency (aliased) theories, whereas the continuous recording (non-aliased) theories are shown as thin black lines.
Green and blue lines indicate $f^{-2}$ and $f^{-4}$ behavior, respectively.
Simulation parameters:   $D=0.46\,\mu$\,m$^2$/s,  $T = 275$\,K,  $m=1$\,ng, and $\fc = 500$\,Hz,  $\fsample = 32,768$\,Hz, $N=131,072$, $\nwin = 32$ Hann windows.
The simulation parameters for the OT case  are those of a 1\,$\mu$m diameter polystyrene sphere held in an optical trap in water at room temperature.
The parameters for Einstein's Brownian motion are the same, except $\kappa = 0$.
For the OU process we increased the density of the sphere roughly 2,000 times, which is not a physically realistic scenario but allows us to plot all power spectra with the same axes. 
}
\end{figure}

\begin{figure}
\includegraphics[width=\linewidth]{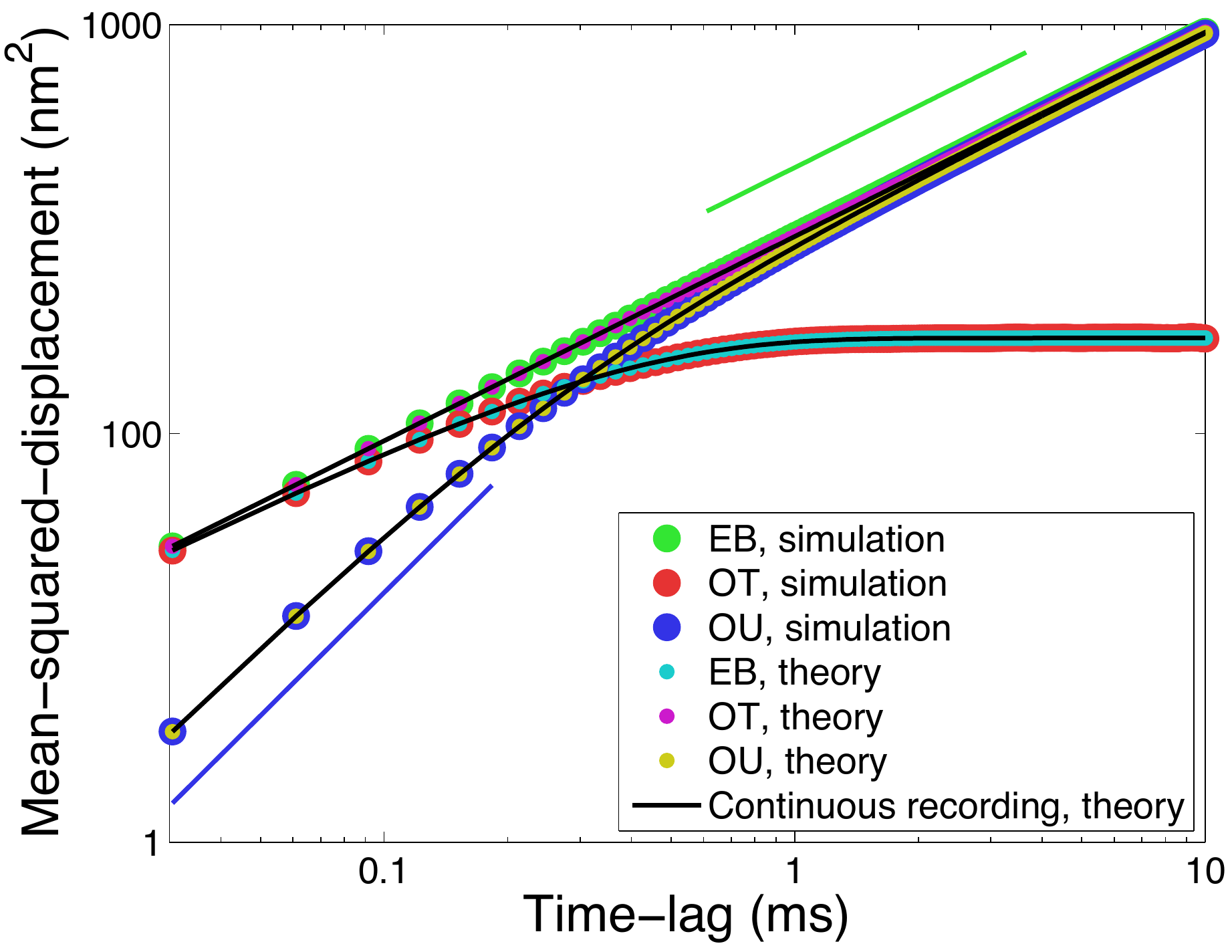}
\caption[]{\label{fig:OT_MSD}
Mean-squared-displacement for the Ornstein-Uhlenbeck, optical-trap, and Brownian-motion limits described in Sec.~\ref{sec:OU}, \ref{sec:OT}, and \ref{sec:BM}.
Simulation settings and legends are the same as in Fig.~\ref{fig:OT_PSDs}, except green/blue lines show slopes of one/two.
}
\end{figure}

Figures~\ref{fig:OT_PSDs}~and~\ref{fig:OT_MSD} show the power spectra, respectively mean-squared-displacements obtained in numerical simulations of free diffusion and trapped diffusion, as well as the graphs of the corresponding analytical expressions.
At short time-scales, i.e., at high frequencies in Fig.~\ref{fig:OT_PSDs} and for  small time-lags in Fig.~\ref{fig:OT_MSD}, the thermal forces dominate and the Hookean force has not had time to influence the motion through its constant, but weak, confining effect.  That is why the Brownian motion (green) and optical trap (red) data collapse in this regime, whereas the Ornstein-Uhlenbeck process (blue) differ from the two due to inertial effects.
Conversely, at long time scales,  low frequencies in Fig.~\ref{fig:OT_PSDs} and large time-lags in Fig.~\ref{fig:OT_MSD}, inertia plays no role, so the Ornstein-Uhlenbeck process and Einstein's theory for Brownian motion are indistinguishable, whereas the Hookean force has had time to exert its confining effect on the optical trap data.

\section{Nondimensionalization}
Above, we worked with dimension-full equations because they are physically intuitive and allow for dimensional checks of the calculations along the way.
Extra insight can be gained however, by nondimensionalizing the equations.

As an example, in the dynamic equation for the OT case
\beq
\gamma \dot{x} + \kappa x = \Ftherm(t) = \sqrt{2 \kB T \gamma} \, \eta(t)  \e,
\eeq
there are three dimension-full parameters and two dimension-full variables.
Dividing through with one of the parameters there are only two left.  
From these we can form a characteristic time and a characteristic length.
By expressing times and lengths in terms of these, we have removed all parameters from the equation.
In other words, there is only a single unique equation.

What is a natural choice for the characteristic time and length?
By considering the physics, we see that switching off the driving-force will result in a solution to the dynamics that is an exponentially decaying distance from $x(t)$ to $x=0$.  The characteristic time, $\gamma / \kappa$, for this decay is the characteristic time for the dynamics and it makes sense to measure time in units of this time. 
With the driving-force on, a statistical equilibrium establishes itself.  
In this equilibrium the distribution of positions has a certain width that is temperature-dependent because the driving-force is thermal.
This width $\sqrt{\la x^2 \ra}$, is a natural unit in which to measure distances and gives us a position variable that is dimensionless.

The basic idea then, is to express the independent variable $t$, and dependent variable $x$ in terms of dimensionless ones and then use the freedom we have to choose parameters to make as many pre-factors as possible equal to unity \cite{Buckingham1914}. 
Above we used a physical approach.
In a more mathematical approach the recipe is to write  $x = x_c \chi$ and $t=t_c\theta$ where $x_c$ and $t_c$ are dimension-full ``characteristic'' quantities, whereas $\chi$ and $\theta$ are the new dimensionless dependent and independent variables respectively.
Dividing by $\gamma$ we first get
\beq
\frac{x_c}{t_c} \frac{d\chi}{d\theta} + \frac{\kappa}{\gamma}x_c\chi = \frac{1}{\gamma} f_{\rm therm}(\theta)
\eeq
where 
\beq
f_{\rm therm} (\theta) = \Ftherm(t_c\theta) = \sqrt{2\kB T \gamma /t_c} \, \breve{\eta}(\theta)
\eeq
with
\beq
\breve{\eta}(\theta) = \sqrt{t_c} \eta(t_c \theta)
\eeq
and
\beq
\la \breve{\eta}(\theta) \breve{\eta}(\theta') \ra = \delta(\theta-\theta'); \,\,\,\, \la \breve{\eta} \ra = 0 \e.
\eeq
Note, that $\breve{\eta}$ is a dimensionless generalized function of the dimensionless parameter $\theta$, whereas $\eta$ is a generalized function of $t$ and has dimension $1/\sqrt{t}$. 
Next, by choosing
\beq
t_c = \gamma / \kappa \mbox{ and } x_c =  \sqrt{2\kB T/\kappa}
\eeq
we get
\beq
\frac{d\chi}{d\theta} + \chi = \breve{\eta}(\theta)\e,
\eeq
as our dimensionless dynamic equation.
Without having actually solved the original equation we have gained three insights: 
(i) The system has a characteristic time-scale of $\gamma/\kappa$;
(ii) the system has a characteristic length-scale of $\sqrt{2\kB T/\kappa}$ (compare to the MSD);
and (iii) there is only \emph{one} OT-equation, not a whole family for different values of $\gamma$ and $\kappa$.

Since the OU-process for the velocity
\beq
m\dot{v} + \gamma v = \Ftherm(t); \,\,\,\,\, v = \dot{x}
\eeq
is mathematically identical to the OT process for the positions we directly have the dimensionless pendant
\beq
\frac{d\nu}{d\theta} + \nu = \breve{\eta}(\theta)\e,
\eeq
by writing $v=v_c \nu$, $t=t_c\theta$, and choosing
\beq
t_c = m/\gamma \mbox{ and } v_c = \sqrt{2\kB T / m} \e.
\eeq

For completeness, we also give the dimensionless version of the original dynamics.
The dimension-full equation
\beq
m\ddot{x} + \gamma \dot{x} + \kappa x = \Ftherm(t)
\eeq
can be rewritten as the universal oscillator equaition
\beq
\frac{d^2\chi}{d\theta^2} + 2\lambda\frac{d\chi}{d\theta} + \chi = \breve{\eta}(\theta)
\eeq
where $\lambda = \gamma / \sqrt{m \kappa}$ is the damping ratio, $x_c^2 =2\kB T \gamma /  \sqrt{m\kappa^{3}}$, and $t_c^2 = m/\kappa$.  
That is, there are three dimension-full parameters and a family of solutions parametrized by $\lambda$.

To arrive at dimensionless versions of the discrete equations we can proceed to solve the above dimensionless equations as before. 
Alternatively, we can simply take the already known dimension-full results and nondimensionalize them.

\section{Summary and conclusions}
We examined the dampened harmonic oscillator and three of its physical limits: The mass-less case (optical trap), the free  case (the Einstein-Ornstein-Uhlenbeck theory for Brownian motion), and the mass-less free case (Einstein's Brownian motion).
By solving the system's dynamical equations for an arbitrary time-lapse $\Dt$, exact analytical expressions were derived for the changes in position and velocity during such a time-lapse.
With these expressions, exact simulations of the dynamics are then possible---with an accuracy that is independent of the duration of the time-lapse. In contrast, a numerical simulation, using Euler-integration or similar schemes, is exact only to first or second order in $\Dt$ \cite{Mannella2000}.

We gave exact analytical expressions for power-spectral forms, mean-squared-displacements, and correlation-functions that can be fitted (see \cite{Norrelykke2010} before undertaking a least-squares-fit) to data obtained from time-lapse recording of a system with dynamics similar to the dampened harmonic oscillator or one of its three physical limits described here.  The effect of finite sampling rates (aliasing) 
were also discussed. 

The effect on the power spectrum of velocity estimation from position-data (secant-velocity) was treated for the case of free diffusion of a massive particle. Approximate as well as exact corrective factors and expressions were given.
Finally, we discussed some of the advantages of dimensional analysis and of nondimensionalization of the dynamic equations.
Throughout, we pointed out when power spectral analysis makes sense (bounded process) or not, which of the statistical measures that depend on the sampling-frequency, and which may be described by the simpler continuous-time theory.

\section{Acknowledgments}
S.F.N. gratefully acknowledges financial support from the Carlsberg Foundation and the Lundbeck Foundatiion.


\end{document}